\documentstyle[12pt]{article}

\makeatletter

\input a4.sty
\def\@citex[#1]#2{%
\if@filesw \immediate \write \@auxout {\string \citation {#2}}\fi
\@tempcntb\m@ne \let\@h@ld\relax \def\@citea{}%
\@cite{%
  \@for \@citeb:=#2\do {%
    \@ifundefined {b@\@citeb}%
      {\@h@ld\@citea\@tempcntb\m@ne{\bf ?}%
      \@warning {Citation `\@citeb ' on page \thepage \space undefined}}%
      {\@tempcnta\@tempcntb \advance\@tempcnta\@ne%
      \@tempcntb\number\csname b@\@citeb \endcsname \relax%
      \ifnum\@tempcnta=\@tempcntb 
        \ifx\@h@ld\relax%
          \edef \@h@ld{\@citea\csname b@\@citeb\endcsname}%
        \else%
          \edef\@h@ld{\ifmmode{-}\else--\fi\csname b@\@citeb\endcsname}%
        \fi%
      \else
        \@h@ld\@citea\csname b@\@citeb \endcsname%
        \let\@h@ld\relax%
      \fi}%
    \def\@citea{,\penalty\@highpenalty\,}%
  }\@h@ld
}{#1}}
\@addtoreset{equation}{section}
\def\section{\@startsection {section}{1}{\z@}{-3.5ex plus -1ex minus
 -.2ex}{2.3ex plus .2ex}{\large\bf\centering}}
\def\subsection{\@startsection{subsection}{2}{\z@}{-3.25ex plus -1ex minus
 -.2ex}{1.5ex plus .2ex}{\sc}}
\gdef\@publabel{\hfil}
\gdef\@pubdate{\null}
\gdef\@pubnumber{\null}
\gdef\@author{\null}
\gdef\@title{\null}
\gdef\@abstract{\null}
\long\def\pubdate#1{\gdef\@pubdate{#1}}
\long\def\pubnumber#1{\gdef\@pubnumber{#1}}
\long\def\publabel#1{\gdef\@publabel{#1}}
\long\def\author#1{\gdef\@author{#1}}
\long\def\title#1{\gdef\@title{#1}}
\long\def\abstract#1{\gdef\@abstract{#1}}
\def\titlerelax{
\let\maketitle\relax
\let\settitleparameters\relax
\let\consolidatetitle\relax
\let\inittitlepage\relax
\let\finishtitlepage\relax
\let\titlepagecontents\relax
\let\multithanks\relax
\let\titlebaselines\relax
\let\@makepub\relax
\let\@maketitle\relax
\let\@makeauthor\relax
\let\@makeabstract\relax
\let\@maketitlenote\relax
\let\thanks\relax
\let\titlerelax\relax}
\def\titleclean
{\gdef\@titlenote{}
\gdef\@abstract{}
\gdef\@author{}
\gdef\@title{}
\gdef\@pubdate{}\gdef\@pubnumber{}\gdef\@publabel{}
\gdef\@dpublabel{}
}
\def\@makepub{\vbox to \z@{\hbox to \textwidth{\hfill
\@publabel \hfill
\llap{\parbox[t]{0.33\textwidth}{\raggedleft\@pubnumber}}}%
\vss}}
\def\@maketitle{\vskip 60pt \begin{center}
 {\LARGE \@title \par}
 \end{center}}
\def\@makeauthor{{%
\def\and{\smallskip {\normalsize \rm and\smallskip }}
\def\And{\medskip {\normalsize \rm and\\}\medskip}
\long\def\address##1{{\def\and{\\and\\}\medskip
				{\small \it \\##1\\}
}}
{\centering
 \vskip 3em
 \large \lineskip .75em
 \@author}
 \par}}
\def\@makedate{\vskip 1.5em
 {\raggedright \small \noindent\@pubdate \par}}
\def\@makeabstract{\vskip 1.5em
{\small
\begin{center}
{\bf ABSTRACT\vspace{-.5em}\vspace{0pt}}
\end{center}
\quotation \@abstract \endquotation}}
\def\maketitle{\titlepage
\let\footnotesize\small \setcounter{page}{0}
\@makepub
\vfil
\@maketitle
\@makeauthor
\vfil
\@makeabstract
\@thanks
\vfil
\@makedate
\if@restonecol\twocolumn \else \eject \fi
\titlerelax \titleclean
\setcounter{footnote}{0}
}

 \font\tenmsa=msam10 scaled \magstep1
 \font\sevenmsa=msam8
 \font\fivemsa=msam6
 \font\tenmsb=msbm10 scaled \magstep1
 \font\sevenmsb=msbm8
 \font\fivemsb=msbm6

\newfam\msafam
\newfam\msbfam
\textfont\msafam=\tenmsa  \scriptfont\msafam=\sevenmsa
  \scriptscriptfont\msafam=\fivemsa
\textfont\msbfam=\tenmsb  \scriptfont\msbfam=\sevenmsb
  \scriptscriptfont\msbfam=\fivemsb

\def\Bbb{\ifmmode\let\next\Bbb@\else
 \def\next{\errmessage{Use \string\Bbb\space only in math mode}}\fi\next}
\def\Bbb@#1{{\Bbb@@{#1}}}
\def\Bbb@@#1{\fam\msbfam#1}
\def\cW{{\cal W}}
\def\cev#1{\langle #1 \vert}
\def\D#1{\frac{d#1}{2\pi i}}
\def\en{\end{equation}}
\def\eq{\begin{equation}}

\def\vac{\vec 0}
\def\vec#1{\vert #1 \rangle}

\def\WA{\mathop{\it WA}\nolimits}
\def\wa#1{$\WA_#1$}

\def\thebibliography#1{\section*{References\@mkboth
 {REFERENCES}{REFERENCES}}\list
 {[\arabic{enumi}]}{\settowidth\labelwidth{[#1]}\leftmargin\labelwidth
 \advance\leftmargin\labelsep
 \usecounter{enumi}}
 \def\newblock{\hskip .11em plus .33em minus .07em}
 \sloppy\clubpenalty4000\widowpenalty4000
 \itemsep=0pt
 \small
 \sfcode`\.=1000\relax}

\makeatother

\begin{document}

\pubnumber{DAMTP--94--26\\ hep-th/9403163}
\pubdate{24 March, 1994 }
\title{Fusion in the $W_3$ algebra}

\author{
G.\ M.\ T.\ Watts%
\thanks{E-mail address: \tt g.m.t.watts@damtp.cambridge.ac.uk}%
\address{St.\ John's College, St.\ John's Street, Cambridge, CB2 1TP,
U.\ K.
\\ and \\
DAMTP, University of Cambridge, Silver Street, Cambridge, CB3 9EW,
U.\ K.}%
}

\abstract{
We develop the notions of fusion for representations of the
\wa2 algebra along the lines of Feigin and Fuchs.
We present some explicit calculations for a \wa2 minimal model.
}

\maketitle

\openup1\jot

\section{Introduction}

The concept of fusion is central in the application of algebraic
techniques in two-dimensional conformal field theory. In conformal
field theory one supposes the presence of an infinite dimensional
symmetry algebra, and local fields which transform under the algebra.
The local fields are operator valued distributions, and it is taken as
an axiom that the product of two fields may be written as a sum of
fields; this is the operator product expansion. In particular there
is a particular class of fields called primary fields, and in its
simplest form the fusion algebra describes which irreducible
representations $\rho_k$ of
the symmetry algebra can occur in the operator product of two primary
fields, which we write symbolically as
\eq
\Phi_i \times \Phi_j \mapsto N_{ij}{}^k \rho_k
\label{eq.fuse}
\en
where $N_{ij}{}^k$ are the Verlinde fusion algebra coefficients, and
are integers or infinite. For algebras with a non-zero central
extension  the operator product of two fields cannot simply correspond
to the tensor product of two highest weight representations, as in the
former case the value of the central charge is unchanged, whereas it
adds under tensor product.

The simplest non-trivial algebra with which one must deal is the
Virasoro algebra. Belavin, Polyakov and Zamolodchikov showed how null
vectors of the Virasoro algebra affected allowed fusions \cite{BPZ},
but Feigin and Fuchs were the first to translate their ideas into
mathematical language and were able to prove the conjectured fusion
rules of the Virasoro minimal models, as well as providing an
algebraic definition of a minimal model in terms of a quasi-finite
representation \cite{FeFu3}.
A standard treatment would be to consider the detailed structure of
the representation $\rho_i$ in eqn.\ (\ref{eq.fuse}) and find the
constraints on the allowed representations $\rho_k$. Feigin and Fuchs'
method differs from a standard approach in that they  consider instead
the representation $\rho_k$, and derive constraints on the pairs of
fields $\Phi_i$, $\Phi_j$ which can couple to it. Their method also
naturally extends to any number of fields
\eq
\Phi_1 \times \ldots \times \Phi_n \to \rho_0
\label{eq.1n0}
\en
It is the aim of this paper to show how to extend these results to
W-algebras, and in particular the $W_3$ or \wa2 algebra introduced by
Zamolodchikov in \cite{Zamo1}. We shall
show how these can be adapted to the \wa2 algebra, for 3-pt and
$n$-pt functions, we provide a couple of
examples and conclude with suggestions how the main conjectures in
this paper may be proven.

\newpage

\section{Quantum $W\!A_2$ algebra and its representations}
\label{sec.qw}

We take the quantum \wa2 algebra to have generators $L_m$, $Q_m$ with
relations
\begin{eqnarray}
{}~[ L_m, L_n ] &=& \frac c{12}m(m^2-1)\delta_{m+n,0} + (m-n)L_{m+n}
	\nonumber\\
{}~[ L_m, Q_n ] &=& (2m-n)Q_{m+n} \nonumber\\
{}~[ Q_m, Q_n ] &=&
\frac{(22 + 5c)}{48}\frac{c}{3\cdot 5!} (m^2-4)(m^2-1)m\delta_{m+n}
\\
&& + \frac{1}{3}(m-n)\Lambda_{m+n} +
\frac{(22 + 5c)}{48}\frac{(m-n)}{30}(2m^2-mn+2n^2-8)L_{m+n}
\,,
\nonumber
\end{eqnarray}
where
\eq
\Lambda_m = \sum_{p>-2} L_{m-p}L_p
	+   \sum_{p\leq -2} L_p L_{m-p}
	-   \frac3{10}(m+2)(m+3)L_m
,
\en
and $c$ is a central element.

The representation theory of the $W_3$ algebra can be developed in
analogy with that of the Virasoro algebra. A $W_3$ highest weight
vector $\vec{h,q}$ satisfies
\eq
L_m\vec{h,q}=\delta_{m,0}h\vec{h,q}\;,\;
Q_m\vec{h,q}=\delta_{m,0}q\vec{h,q}\;,\;
m \ge 0\,.
\en
The Verma module  $V_{h,q,c}$ of the $W_3$-algebra is spanned by states
of the form
\eq
L_{i_1} \ldots L_{i_j} Q_{k_1} \ldots Q_{k_l} \vec{h,q},\;\;
i_m \leq i_{m+1} \leq -1,\;
k_m \leq k_{m+1} \leq -1,
\label{eq.vbasis}
\en
and by the usual abuse of notation the central element $c$ takes the
value $c$.
If the Verma module is reducible, then the irreducible representation
$L_{h,q,c}$ is the quotient of the Verma module by its maximal
invariant submodule.

We can parametrise the weights
of a W-highest weight vector as follows \cite{FZam4},
\eq
h = \frac{1}{3}( x^2 + xy + y^2 - 3 a^2)
\;,\;
q = \frac{1}{27}(x-y)(2x+y)(x+2y)
\,,
\label{eq.weig}
\en
where we define $a,\alpha_\pm$ by
\eq
c= 2 - 24 a^2 \;,\; \alpha_\pm^2 - \alpha_\pm a - 1 = 0
\,.
\label{eq.cval}
\en
The condition that $V_{h,q,c}$ has a null vector with eigenvalues
$h',q'$ is that we can find some $x, y$ such that $h, q$ are given
by eqns.\ (\ref{eq.weig}) and $x$ satisfies
\eq
x = r\alpha_+  + s\alpha_-,
\;\; r,s\in \Bbb N,\;\; rs>0,
\en
in which case $h',q'$ are given by eqns.\ (\ref{eq.weig}) with
$x' = x - 2 r\alpha_+$, $y'= y + r\alpha_+$.
If
\eq
x=r\alpha_+ + s\alpha_-\;\;,\;
y=t\alpha_+ + u\alpha_-\;\;;\;
r,t,s,u\in\Bbb N^+
\label{eq.hq}
\en
then there are two independent null states in $V_{h,q,c}$ and we call
such a representation doubly-degenerate. We write $h$ and $q$ as
$h[rt;su], q[rt,su]$ and the highest weight state as $\vec{rt;su}$, or
simply denote the representation by $[rt;su]$.

The $W_3$ minimal models are those which have
\eq
\alpha_+ = \sqrt{p/q}\,,\;
p,q\in\Bbb N\,,\; p,q \hbox{ coprime, }
c = c(p,q)= 50 - 24 \frac pq - 24 \frac qp
\label{eq.minmo}
\en
and the fields in these models are of the form $[rt;su]$ with
$0<r,s,t,u$, $r+t<q$, $s+u<p$ (see \cite{Fluk}).
These representations have three independent null vectors in
$V_{h,q,c}$. Those minimal models with $p=m+1$, $q=m$,
$m\geq 3$ are unitary since they can be constructed in the explicitly
unitary coset construction \cite{BBSS}.

For each state $\vec\psi$ in a highest weight representation, we can
define a field, $\psi(z)$ such that $\psi(0)\vac = \vec\psi$.
For each operator $X_m$ where $X=L$ or $Q$ we can define the field
$\hat X_m \psi(z)$ by
$\hat X_m \psi(0)\vac = X_m \vec\psi $. Then it is possible to write
the commutation relations of $L_m$ and $Q_m$ with an arbitrary field
as a sum,
\begin{eqnarray}
{}~[ L_m,\psi] &=& \sum_{j=-1}^\infty z^{m-j}\pmatrix{m+1 \cr j+1}
			\hat L_j \psi
\label{eq.lp}
\\
{}~[ Q_m,\psi] &=& \sum_{j=-2}^\infty z^{m-j}\pmatrix{m+2 \cr j+2}
			\hat Q_j \psi
\label{eq.qp}
\end{eqnarray}
With a  primary field $\Phi_{h,q}(z)$, $L_m$ and $Q_m$ have
commutation relations
\begin{eqnarray}
{}~[L_m, \Phi_{h,q}(z) ] &=&
\left(h (m+1)z^{m} + z^{m+1}\partial\right)  \Phi_{h,q}(z) \\
{}~[Q_m, \Phi_{h,q}(z) ] &=&
\left( \frac q2(m+2)(m+1)z^m + (m+2)z^{m+1}\hat Q_{-1}
	+ z^{m+2}\hat Q_{-2} \right) \Phi_{h,q}(z)
\;,
\nonumber
\end{eqnarray}
Since $\hat L_{-1}\psi(z) = \partial\psi(z)$, this results in  a
representation of the
Virasoro algebra on the modes of $\psi(z)$, whereas
for the modes $Q_m$  this is not possible, as the commutator
(\ref{eq.qp}) includes the new fields
$ \hat Q_{-1}\psi(z)$ and $\hat Q_{-2} \psi(z)$.
This fact is responsible for many of the difficulties in the theory of
the \wa2 algebra.

The approach of Feigin and Fuchs is to consider the correlation
functions of the form (\ref{eq.1n0}) as a map $\varphi$ from the
irreducible representation $L_{h,q,c}$,
\eq
\varphi \;:\; \vec\psi \mapsto
\cev {h^\infty, q^\infty}
\prod_{i=1}^n \Phi_{h^i, q^i}(w_i)
\vec\psi
\en

Although the modes $L_m$ and $Q_m$ do not have nice commutation
relations with primary fields,
as for the Virasoro case \cite{FeFu3} one can define combinations of
these modes which have much simpler commutation relations.
By cancelling all the poles in the operators product of $L(z)$ or
$Q(z)$ with a primary field, we can arrange that certain operators $e_m$
and $f_m$ commute with a field, and
by cancelling all but the leading pole we can pick out the weights of
the fields.
If we wish to consider correlation functions of the form
\eq
\cev {h^\infty, q^\infty}
\prod_{i=1}^n \Phi_{h^i, q^i}(w_i)
\vec\psi\;,\;\; w_i \neq 0
\label{eq.corf}
\en
for arbitrary states $\vec\psi$, then it is sensible to consider the
operators
\begin{eqnarray}
e_m &=& \oint_0
	\left( \prod_{i=1}^n \left( \frac{ z-w_i}{z w_i} \right)^2
	\right)
	z^{m+1} L(z) \D z
\,\,\,=\,\,\,
	L_{-2n+m} \,+\, \ldots \,+\, L_{m} \prod_i w_i^{-2}
\nonumber\\
f_m 	&=& \oint_0
	\left( \prod_{i=1}^n \left( \frac{ z-w_i}{z w_i} \right)^3
	\right)
	z^{m+2} Q(z) \D z
\,\,\,=\,\,\,
	Q_{-3n+m} \,+\, \ldots \,+\, Q_{m} \prod_i w_i^{-3}
{}~~~~~~~~~~
\label{eq.emfm}
\end{eqnarray}
The operators $e_m$ and $f_m$ clearly depend on the points $w_i$ but
we suppress the dependence if it is clear from context.
If the points $w_i$ are finite and distinct, then we can also consider
the operators
\begin{eqnarray}
e_0^{(w_j)} &=& - \oint
	\prod_{i\neq j}
	\left( \frac{ (z- w_i)w_j}{(w_j - w_i)z} \right)^2
	\frac{ (z- w_j) w_j}{z}
	L(z) \D z
\nonumber\\
f_0^{(w_j)} &=& - \oint
	\prod_{i\neq j}
	\left( \frac{ (z- w_i)w_j}{(w_j - w_i)z} \right)^3
	\left( \frac{ (z- w_j) w_j}{z} \right)^2
	Q(z) \D z
\label{eq.e0f0}
\end{eqnarray}
For these operators
\begin{eqnarray}
&& \cev{h^\infty, q^\infty} \;
	\prod_i \Phi_{h_i,q_i}(w_i) \,(e_m  - \delta_{m,0}h^\infty )
	= 0 \;,\;\; m \leq 0 \nonumber\\
&& \cev{h^\infty, q^\infty} \;
	\prod_i \Phi_{h_i,q_i}(w_i) \,(\,  e_0^{(w_j)} - h_j \,)
	= 0 \nonumber\\
&& \cev{h^\infty, q^\infty} \;
	\prod_{i} \Phi_{h_i,q_i}(w_i) \,(f_m  - \delta_{m,0} q^\infty)
	= 0 \;,\;\; m \leq 0 \nonumber\\
&& \cev{h^\infty, q^\infty} \;
	\prod_{i} \Phi_{h_i,q_i}(w_i) \,(\,f_0^{(w_j)} - q_j\,)
	= 0
\label{eq.tpf}
\end{eqnarray}
Let us define for $w_i \neq 0$
\eq
\cW_<( w_1,\ldots w_n) = {\rm span}\, (e_m, f_m, m<0)
\;,
\en
and in the case $w_i$ finite and distinct,
\eq
\cW^0( w_1,\ldots w_n
) = {\rm span}\, (e_0^{(w_1)}, f_0^{(w_1)},\ldots;
e_0^{(\infty)}, f_0^{(\infty)} )
\en
For any highest weight representation space $N$, it is easy to see by
direct calculation of their commutators that the elements
of $\cW^0$ act as an Abelian algebra on $N / \cW_<  N$.

The space of maps $\varphi$ which satisfy (\ref{eq.tpf}) is given as
$L\,/\, \cW_< L$,  and the space of maps $\varphi_{\{h^i,q^i\}}$ with
fixed values of weights is given as the quotient of $L\,/\, \cW_< L$
by the relations $\{e_0^{(w^i)} -h^i,f_0^{(w^i)} -q^i,\}$.  The
structure of $L\,/\, \cW_< L$ as a representation of $\cW^0$ may lead
to restrictions on the allowed  values of $\{h^i,q^i\}$. In the worst
case there are no restrictions on the weights of the fields and the
space of maps $\varphi_{\{h^i,q^i\}}$ is infinite dimensional for each
choice of weights. In the best case, the space of maps $\varphi$ is
finite dimensional  and hence there is only a finite set of allowed
weights   $\{h^i,q^i\}$ which give non-zero correlation functions with
the irreducible representation $L_{h,q,c}$. If the points $w_i$ are
distinct then in this case we expect $\cW^0$ to be  diagonaliseable
and the eigenvalues of $\{e_0^{(w_i)}, f_0^{(w_i)}\}$ give the allowed
weights.

\leftline{\bf Proposition:}

\hang
$\dim L \,/\, \cW_<(w_1\ldots w_n) L $ is independent of the points
$w_i$ provided these are all non-zero. In particular the dimension
does not change if $w_i$ are coincident or infinite.

\noindent
The Verma module $V_{h,q,c}$ has a basis (\ref{eq.vbasis}), but there
is another basis
\eq
e_{i_1}\ldots e_{i_p}
f_{j_1}\ldots f_{j_q}
L_{k_1} \ldots L_{k_r}
Q_{l_1} \ldots Q_{l_s} \vec{h,q}\,,
\label{eq.efbasis}
\en
\[
i_m \leq i_{m+1} \leq -1\,,\;
j_m \leq j_{m+1} \leq -1\,,\;
-2n \leq k_m \leq k_{m+1} \leq -1\,,\;
-3n \leq l_m \leq l_{m+1} \leq -1\,.
\]
Since $e_m(w_1,\ldots w_n) = L_{-2n+m} + \ldots$ then the bases
(\ref{eq.efbasis}) for different sets $\{w_i\}$ are related by upper
triangular matrices with ones on the diagonal.  We can use the
relations in  the irreducible representation $L_{h,q,c}$ to find a
basis  which consists of a subset of the vectors (\ref{eq.efbasis}).
Hence if the vectors $u_i$ are independent in $L \,/\, \cW_<(w_i) L$
they are also independent in $L \,/\, \cW_<(w'_i) L$ and so the
dimension of this space is independent of the points $\{w_i\}$.

If we are only interested in the dimensionality of $L \,/\, \cW_< L$
then we can choose the $w_i$ all infinite and restrict attention to
the simpler space $L\,/\, \cW_n L$ where
\begin{eqnarray}
\cW_n &=& \cW_<(\infty\ldots\infty) \nonumber\\
	&=&
{\rm span}\, ( L_{-2n-1}, \ldots; Q_{-3n-1}, \ldots )
\end{eqnarray}
Whereas for the $w_i$ finite and distinct $L \,/\, \cW_< L$ carries a
representation of $\Bbb C^{2n+2}$ generated by $\cW^0$, the space
$L \,/\, \cW_n L$ carries a representation of $\Bbb C^{2n+2}$
generated by
\eq
{\rm span}\, ( L_{-n}, \ldots L_{-2n}; Q_{-2n} ,\ldots Q_{-3n})
\en

\subsection{Three-point functions}

Let us consider a three point function
\eq
\cev{h_1,q_1} \Phi_{h_2,q_2}(w) \vec\psi
\label{eq.tpf2}
\en
{}From the preceding discussion, we need only consider the space
\eq
L_{h,q,c} \,/\, \cW_<(1) L_{h,q,c}
\en
and if we are interested only in the dimensionality of this space we
can restrict attention to the somewhat simpler space
\eq
\tilde L_{h,q,c}
=
L_{h,q,c} \,/\, \cW_1 L_{h,q,c}
\en
$\tilde L_{h,q,c}$ is a quotient of the space $\tilde V_{h,q,c}$,
\eq
\tilde V_{h,q,c}
=
V_{h,q,c} \,/\, \cW_1 V_{h,q,c}
\en
where $\tilde V_{h,q,c}$ has a canonical basis
\eq
U(\cW_-) \vec{h,q,c}
\;,\;\;
\cW_- = {\rm span} \left(L_{-1},L_{-2},Q_{-1},Q_{-2},Q_{-3} \right)
\en
If $M_{h,q,c}$, the maximal invariant submodule of $V_{h,q,c}$, is
generated by a finite set of highest weight null states $N_\imath$
then
\eq
\tilde L = \tilde V \,/\, \tilde M
\;,\;\;
\tilde M = U(\cW_-) {\rm span} \left( N_\imath \right)
\en
As mentioned above, the modes $\{L_{-1}, L_{-2}, Q_{-2}, Q_{-3}\}$ act
as an Abelian algebra on $\tilde L$, but we can also consider the
whole of $\cW_-$, which (acting on $\tilde L$) has relations
\eq
\begin{array}{rclrcl}
{}~[Q_{-1},L_{-1}] &=& Q_{-2}
	&
{}~[Q_{-1},L_{-2}] &=& 3 \,Q_{-3}
	\\[2mm]
{}~[Q_{-1},Q_{-2}]
	&\sim&
	\frac 23 L_{-2}L_{-1}
	&
{}~[Q_{-1},Q_{-3}]
	&\sim&
	\frac 23 L_{-2}L_{-2}
\end{array}
\en 
We can represent this in terms of a differential polynomial ring, with
generators
\[
 L_{-1}=x, \;
L_{-2}=y, \;
Q_{-2}=z, \;
Q_{-3}=u, \;
\]
\eq
Q_{-1}=
\!D\!=
 z\frac{\partial}{\partial x}
+ 3u\frac{\partial}{\partial y}
+ \frac 23 xy\frac{\partial}{\partial z}
+ \frac 23 y^2\frac{\partial}{\partial u}
\en
$D$ has a non-trivial kernel, containing, amongst other things,
$ - 27 u^2 + 4 y^3 $ and $  - 9 u x z + x^2 y^2 + 3 y z^2 $.
This is particularly useful in the vacuum representation as we need
only consider the action of $D$ on the null highest-weight states to
find all the restrictions on $\tilde L_{0,0,c}$.

\subsection{Classes of \wa2 representations}

There is not yet a complete classification of \wa2 Verma module
structures, and so we can present only some partial results. In
ref.\ \cite{BWat4} we considered four classes of \wa2  algebra
representations. Of these the classes 2c(i) and 1(c) of ref.\
\cite{Watt1a} are of special interest, as `quasi-rational' and
`quasi-finite' representations respectively.

We say an irreducible highest weight  representation is quasi-finite
if
\eq
\dim \left( L_{h,q,c} \,/\, \cW_<(w_1, \ldots, w_n) L_{h,q,c} \right)
	< \infty
\en
for any set of non-zero points $w_i$. We have

\leftline{\bf Conjecture}
\hang
A representation $L_{h,q,c}$ is quasi-finite if and only if it is a
minimal model representation.
If it is quasi-finite, then for any set of distinct points
$\{ w_i\}$, $\cW^0(\{ w_i\})$ is diagonaliseable on
$  L_{h,q,c} \,/\, \cW_<(\{ w_i\}) L_{h,q,c} $.

\noindent
The term quasi-rational has been used by Nahm in \cite{Nahm4} for
those representations for which in eqn.\ (\ref{eq.tpf2}) for given
irreducible representation $\rho_1$ there are only a finite number of
allowed representations $\rho_2$. In the context of the \wa2 models,
if we fix $\{h_1,q_1\}$ then  the number of allowed values $d$ of
$\{h_2,q_2\}$ is given by
\begin{eqnarray}
d &=&
	\dim
	L_{h,q,c} \,/\,
	\left\langle
	e_0^{(\infty)},f_0^{(\infty)}, \cW_<(1)
	\right\rangle
	L_{h,q,c}
\nonumber\\
 &=&
	\dim
	L_{h,q,c} \,/\,
	\left\langle
	L_{-2},L_{-3},\ldots; Q_{-3}, Q_{-4} \ldots
	\right\rangle
L_{h,q,c}
\label{eq.qr}
\end{eqnarray}
In Nahm's terminology, the representation is quasi-rational if $d$ is
finite, and eqn.\ (\ref{eq.qr}) is his requirement of quasi-rationality.
In \cite{BWat4} we conjectured that the doubly degenerate
representations are quasi-rational.

At the minimal values for $c$, the quasi-rational representations
acquire a third independent null vector. This will reduce the allowed
fusions of the form (\ref{eq.tpf2}) to a finite set of pairs
$\{h_1,q_1\}$, $\{h_2,q_2\}$.
The restrictions this imposes on the representations which arise in
the minimal models can be derived from the vacuum representation
alone, as Feigin et al.\ described in \cite{FNOo1}.  The fusion rules
of the other $L_{h,q,c}$ can be derived from the structure of $\tilde
L_{h,q,c}$.

We now present two simple examples, the vacuum representation and the
doubly-degenerate representation [11;12], for generic $c$-values, and
then in the case of the minimal model $c=c(7,3)=-114/7$.

\subsection{The vacuum representation}
\label{sec.vac}

The first  representation in which we might be interested is the
vacuum representation. This has weights $h=q=0$ and has null vectors
\[
L_{-1}\vac\,,\; Q_{-1}\vac\,,\; Q_{-2}\vac
\]
for all $c$ values.
As a result, $L_{0,0,c}$  must factor through the space
with basis
\eq
e_{i_1}\ldots e_{i_p}
f_{j_1}\ldots f_{j_q}
L_{-2}^a Q_{-3}^b
\vec{h,q}\,,
\label{eq.efbasis2}
\en
\[
i_m \leq i_{m+1} \leq -1\,,\;
j_m \leq j_{m+1} \leq -1\,.
\]
If $w\neq\infty$, then we can define $e_0^{(\infty)}$, $e_0^{(w)}$,
$f_0^{(\infty)}$ and $f_0^{(w)}$
and it is straightforward to see that
$e_0^{(\infty)} = e_0^{(w)}$
and
$f_0^{(\infty)} = f_0^{(w)}$
on the space $\tilde L_{0,0,c} $
so that the only possible fusions with the vacuum sector are of the
form
\[
\Phi_{h,q} \times \Phi_{h,q} \to L_{0,0,c}\,,\;
\]
so from eqn.\ (\ref{eq.qr})
the vacuum representation is quasi-rational.
For generic $c$ values, $\tilde L_{0,0,c}$ is infinite dimensional, as
we can see by looking at the restriction of the Shapovalov form
(inner product matrix) to
the space spanned by the states (\ref{eq.efbasis2}). The determinant
of this form is non-zero, as the leading contributions come from the
diagonal, and so $\tilde L_{0,0,c}$ is reducible for a countable set
of $c$ values only.

If there is another independent null vector in the vacuum sector, then
there  may be a restriction on the allowed values of $h$ and $q$ in a
field theory. This is expected to be the case for the minimal models,
for which we expect that the number of allowed representations is
given by
\eq
N
=
\sum_{i,j} N_{ij}{}^0
=
\dim L_{0,0,c} \,/\, \cW_1 L_{0,0,c}\,.
\en
We present a calculation of  the space
$L_{0,0,c} \,/\, \cW_<(1) L_{0,0,c}$
for $c=-114/7$ in section \ref{sec.1147}.

\subsection{The representation [11;12]}

This representation has $h$ and $q$ as given in
(\ref{eq.weig},\ref{eq.hq}), with $\alpha = \alpha_+$,
\[
h(11;12) = \frac 4{3\alpha^2}-1 \;\;,\;
q(11;12) = \frac{(5-3\alpha^2)(4-3\alpha^2)}{27\alpha^3}
\,.
\]
There are the following null vectors in the  Verma module
$V_{h[11;12], q[11;12], c}$ for all $c$ values
\begin{eqnarray}
\vec{N_1} &=&
	\left( Q_{-1} + (\frac \alpha2 - \frac 5{6\alpha})L_{-1}
	\right) \vec{11;12}  \nonumber\\
\vec{N_2} &=&
	\left( Q_{-2} + \frac2{3\alpha}L_{-2} - \alpha L_{-1}^2
	\right) \vec{11;12} \nonumber\\
\vec{N_3} &=&
	\left( Q_{-3} - \alpha^3 L_{-1}^3 + (\frac 1{6\alpha} +
	\frac\alpha2) L_{-3} + \alpha L_{-2}L_{-1} \right)
\vec{11;12}
\,.
\label{eq.three}
\end{eqnarray}
The state $\vec{N_3}$ is a descendant of $\vec{N_1}$ and $\vec{N_2}$.
As a result,  $\tilde L_{h[11;12], q[11;12],c}$
factors through the space
with basis
\[
L_{-2}^m L_{-1}^n \vec{11;12}
\]
For generic $c$ values, there are no more identities as we can again
see by looking at the $c\to\infty$  limit.
On $\tilde L$ we see $e_0^{(\infty)}$ and $e_0^{(1)}$ may be taken as
independent, and
from the null vectors (\ref{eq.three})
$f_0^{(\infty)}$ and
$f_0^{(1)}$ are given as
\begin{eqnarray}
f_0^{(\infty)}  \!&\!\!\!\!=\!\!\!\!&\!
\frac{1}{27\alpha^3}
(9 \alpha^4 (e_0^{(\infty)})^2 \!- 18 \alpha^4 e_0^{(\infty)}e_0^{(1)}
	\!+ 9 \alpha^4 (e_0^{(1)})^2 - 9 \alpha^4 - 3 \alpha^2
	e_0^{(\infty)} \!- 6 \alpha^2 e_0^{(1)} + 18 \alpha^2 - 8)
\nonumber\\
&&\times
(3 \alpha^2 e_0^{(\infty)} - 3 \alpha^2 e_0^{(1)} + 1)
\nonumber\\
f_0^{(1)}	\!&\!\!\!\!=\!\!\!\!&\!
\frac{1}{27 \alpha^3}
(9 \alpha^4 (e_0^{(\infty)})^2 \!- 18 \alpha^4 e_0^{(\infty)}e_0^{(1)}
	\!+ 9 \alpha^4 (e_0^{(1)})^2 - 9 \alpha^4 - 6 \alpha^2
	e_0^{(\infty)} \!- 3 \alpha^2 e_0^{(1)} + 18 \alpha^2 - 8)
\nonumber\\
&&\times
 (3 \alpha^2 e_0^{(\infty)} - 3
\alpha^2 e_0^{(1)} - 1)
\,.
\label{eq.f00}
\end{eqnarray}
We find that $\tilde L$ is equivalent to
$\Bbb C[ e_0^{(1)}, e_0^{(\infty)},  f_0^{(1)}, f_0^{(\infty)} ]$
modulo the relations (\ref{eq.f00}), so that the representation
[11;12] is quasi-rational.
We shall again consider this representation and the restrictions which
arise from the extra null vector in the minimal model $c=-114/7$ in
section \ref{sec.1147}.

\section{The model $ c = -114 / 7 $}
\label{sec.1147}
\def\om{\omega}

The minimal models of the \wa2 algebra are parameterised by coprime
integers $p,q$ greater than 2. There are at least two series of special
interest,
$(p,q)= (m,m+1)$ and \break
$(p,q)=(3,q)$. The first is the unitary series,
and the second is a non-unitary series. For the
Virasoro algebra the corresponding non-unitary series of models
$c= c_{Vir}(2,q)$
lead to relations with Gordon identities \cite{FNOo1} and the fusion
rings and representations have special properties. There is every
reason to believe that the $(3,q)$ series of the \wa2 algebra will
also have interesting properties.
Here we shall limit ourselves to the model $(3,7)$ which has 5
representations and central charge $c=-114/7$. We choose
$\alpha = \sqrt{7/3}$, in which case the models representations are
$[11;ab]$ with $1 \leq a$, $1\leq b$, $a+b \leq 6$ with each
representation occurring three times in this list.
We shall focus in particular on the vacuum representation $[11;11]$
and the $[11;12]$ representation, and calculate
the spaces
$L_{h,q,c} \,/\, \cW_1 L_{h,q,c}$
and
$L_{h,q,c} \,/\, \cW_<(1) L_{h,q,c}$
for these two representations.

As a point of notation, we shall use $\equiv$ to denote equivalence in
the irreducible representation $L$ and $\sim$ to denote equivalence in
the space $L \,/\, \cW_1 L$.

\subsection{The vacuum representation}

The vacuum representation has three equivalent parameterisations,
\[
[11;11] \;,\;\; [11;15] \;,\;\; [11;51]
\,,
\]
which implies that there is a null vector at level $5$, which is not a
descendant of those at level 1.
{}From section \ref{sec.vac} we need only consider as a basis of
$ \tilde L_{0,0,c}$ the states
\eq
 L_{-2}^a Q_{-3}^b \vac
\,.
\label{eq.ql}
\en
We can find the explicit form of the additional null state at level 5
from ref.\ \cite{BWat2}.
After reduction modulo the generic vacuum sector relations, we find
the expression
\eq
(
	7 L_{-2} Q_{-3} - 3 Q_{-5}
)\vac
\equiv 0
\label{eq.n15}
\en
which tells us that there is another relation in $L / \cW_1 L$,
\eq
L_{-2} Q_{-3} \vac \sim 0
\,.
\label{eq.n15a}
\en
We can derive extra information from considering the repeated action
of $Q_{-1}$ on the state (\ref{eq.n15}), or alternatively from the
action of $D$ on (\ref{eq.n15a}),
from which we deduce that
\eq
(2 L_{-2}^3 + 9 Q_{-3}^2 )\vac \sim 0
\;,\;\;
L_{-2}^4 \vac \sim 0
\,,
\en
which leaves us with a 5 dimensional basis of $L_{0,0,-114/7} \,/\,
\cW_{2,3}  L_{0,0,-114/7} $, viz.
\[
\vac ,\;
L_{-2} \vac ,\;
Q_{-3} \vac ,\;
L_{-2} L_{-2} \vac ,\;
L_{-2} L_{-2} L_{-2} \vac
\,,
\]
from which we deduce that there are 5 fields in this theory. There can
be no more restrictions on $\tilde L_{0,0,-114/7}$ as  null
states at higher levels clearly cannot reduce the dimension further.

We can consider the extra relations in $L \,/\, \cW_< L$ from the null
vectors at levels 5 and 6, which are
\[
f_0 ( 7e_0 + 3) \vac \sim 0
\,,\;
\left( \, 2 e_0 (7 e_0 + 4) (7 e_0 + 5) + 441 f_0^2\,\right) \vac
\sim 0
\,.
\]
There are five solutions to these equations as expected,
\eq
(e_0, f_0) =
(0,0), (- \frac 47,0 ) , (-\frac 57,0),
(-\frac  37, \pm \frac 2{7\sqrt{21}})
\,.
\en
We can also find a basis of
$ L_{0,0,-114/7}  \,/\, \cW_<(1) L_{0,0,-114/7} $ and diagonalise
the algebra $\cW^0$ on this space. We find that, as expected,
\eq
L / \cW_< L = \oplus_{i=1}^5  \Bbb C v_i
\en
where
\[
e_0^{(1)} v_a = h_a v_a \;,\;\;
f_0^{(1)} v_a = q_a v_a
\,.
\]
Explicitly we can find expressions for representatives of the $v_a$,
\begin{eqnarray}
v_1 &=& (e_0 + 3/7) (e_0 + 4/7)  (e_0 + 5/7)    \vac
\nonumber\\
v_2 &=& e_0 (e_0 + 3/7)  (e_0 + 5/7)    \vac
\nonumber\\
v_3 &=& e_0 (e_0 + 3/7) (e_0 + 4/7)   \vac
\nonumber\\
v_4 &=& f_0 (f_0 + 2/(7\sqrt{21}))   \vac
\nonumber\\
v_5 &=& f_0 (f_0 - 2/(7\sqrt{21}))   \vac
\label{eq.v5}
\end{eqnarray}
for which we find the eigenvalues,
\eq
\begin{array}{@{\extracolsep{2mm}}cccccc}
 & v_1 & v_2 & v_3 & v_4 & v_5 \\[2mm]
\hline\noalign{\medskip}
e_0^{(1)} & 0 & -\frac47 & -\frac57 & -\frac37 & -\frac37
	\\[2mm]
f_0^{(1)} & 0 & 0 & 0& \frac 2{7\sqrt{21}} & -\frac 2{7\sqrt{21}}
\end{array}
\label{eq.v5eig}
\en
which are exactly the allowed representations in the $c= -114/7$
minimal model.
This is very clearly related to the procedure of Feigin et al.\ in
\cite{FNOo1}, and we obtain the same equations from the null vectors
at levels 5 and 6 as they obtain in ref.\ \cite{FNOo1} eqn.\ (4.10).
We may also consider other representations and obtain their fusion
rules in an analogous fashion.

\subsection{ The $[11;12]$ representation}

The $[11;12]$ representation can also be parameterised as the $[11;41]$
and $[11;24]$ representations and there are independent null vectors at
levels 1,2, and 4.
The highest weight state has $L_0$ and $Q_0$ eigenvalues
$ -3/7$ and $\om = 2/(7\sqrt{21})$ respectively.
In $L/\cW_1 L$ the null vectors at levels 1,2 and 3 from
eqn.\ (\ref{eq.three}) which are generic to representations of type
$[11;12]$ imply
\[
Q_{-1} \vec{11;12} \sim -\frac 1{\sqrt{21}} L_{-1} \vec{11;12}
	\,,\;
Q_{-2} \vec{11;12} \sim \frac 1{\sqrt{21}} (7 L_{-1}^2 - 2L_{-2})
	\vec{11;12}
\]
\eq
Q_{-3}  \vec{11;12}  \sim
 \sqrt{\frac 73 } \left( \frac 73 L_{-1}^3 - L_{-2} L_{-1} \right)
\vec{11;12}
\en
which leaves us with a possible basis of
$ L / \cW_{2,3} L$ of the form
\[
L_{-2}^a L_{-1}^b \vec{11;12}
\]
We can now use the independent null vector at level 4,
\[
\left( L_{-2}^2 \,-\, \frac{49}6 L_{-1}^4 \right)\, \vec{11;12} \sim 0
\]
and the repeated action of $Q_{-1}$ on this state to obtain
\[
\left( L_{-2}L_{-1}^3\, - \frac37 L_{-1}^5 \right) \,\vec{11;12}\sim 0
\,,\;
L_{-1}^7\,\vec{11;12} \sim 0
\]
and reduce the
possible basis states to
\eq
\vec{11;12}\,,\;
L_{-2}\vec{11;12}\,,\;
L_{-2} L_{-1}\vec{11;12}\,,\;
L_{-2} L_{-1}^2\vec{11;12}\,,\;
\en
\[
L_{-1}\vec{11;12}\,,\; L_{-1}^2\vec{11;12}\,,\; L_{-1}^3\vec{11;12}\,,\;
L_{-1}^4\vec{11;12}\,,\;
L_{-1}^5\vec{11;12}\,,\;
L_{-1}^6\vec{11;12} \,.
\]

We can now try to find the `fusion basis', that is a basis of
$L\,/\, \cW_<(1) L$.
{}From the null vectors at levels 1
to 7 we obtain equations which lead to a total of 10
solutions for the fusion,
\[
L_{-3/7, \om, -114/7}
 \;/\; \cW_<(1)\, L_{-3/7, \om, -114/7}
= \Bbb C^{10}
\]
with the following eigenvalues of $\cW^0$ on the basis states
\[
\begin{array}{@{\extracolsep{2mm}}ccccccccccc}
 & v_1 & v_2 & v_3 & v_4 & v_5 & v_6 & v_7 & v_8 & v_9 & v_{10} \\[2mm]
\hline\noalign{\medskip}
	e_0^{(\infty)} &
-3/7 & -3/7 & -3/7 & -3/7 & -4/7 & -4/7 & -5/7 & -5/7 & -5/7 & 0
	\\[2mm]
	f_0^{(\infty)} &
-\om & -\om & \om  & \om  & 0    & 0    & 0    & 0    &  0   & 0
	\\[4mm]
	e_0^{(1)} &
-3/7 & -4/7 & -5/7 & 0    & -3/7 & -5/7 & -3/7 & -4/7 & -5/7 & -3/7
	\\[2mm]
	f_0^{(1)} &
\om  &  0   &   0  & 0    &  \om & 0    & -\om & 0    &  0   &  -\om
\end{array}
\]
and so in this case we find that
\[
\dim \, L_{-3/7, \om, -114/7} \,/\,  \cW_1\,
	L_{-3/7, \om, -114/7}
=
\dim \, L_{-3/7, \om, -114/7} \,/\,  \cW_<(1)\,
	L_{-3/7, \om, -114/7}
\,.
\]

\section{Conclusions}

Building on our work in \cite{BWat4}, we have outlined an algebraic
way to
extend  the work of Feigin and Fuchs in ref.\ \cite{FeFu3} to the
\wa2 algebra. Clearly this will extend to all the algebras
$Wg_n$, and probably all the algebras which can be obtained by
generalised Drinfel'd-Sokolov construction.

How could one prove the conjectures we have made here?
The proofs in Feigin and Fuchs relied on their calculation of the
embedding structure of Verma modules of the Virasoro algebra
in \cite{FeFu2}.
The corresponding calculation has not yet been performed for the \wa2
algebra representations, and there are new problems such as the
presence of subsingular vectors, and the fact that the action of $Q_0$
is on many occasions not diagonaliseable on doubly-degenerate
Verma module representations.
The consideration of these problems is work in progress.

Interesting developments which might help are the work on the
structure of finite \hbox{W~algebra} modules by de Vos and van Driel
\cite{dvvd}  and Bajnok's construction of null vectors
of the \wa2 algebra using complex powers of generators \cite{Bajn2}.
Certainly there are some interesting results for the $c=c(3,p)$ models
\cite{Frenkel}.

This method is most suitable for the study of the minimal models, but
for the study of quasi-rational models as suggested by Nahm in
\cite{Nahm4}, it is necessary to consider some more general ideas, and
attempt to construct some form of tensor product of \wa2
representations, as proposed by Gaberdiel in \cite{Gabe1}.

The fusion rules for the representations of W-algebras obtained by
quantum Hamiltonian reduction were obtained in \cite{FKWa1} from the
modular properties of the characters obtained
on the
basis of conjectured resolutions by Wakimoto modules. It would be nice
if one could obtain a direct connection with this work.

\leftline{\bf Acknowledgements}

I would especially like to thank
P.\ Bowcock for his collaboration at an early stage and for the
insights into \wa2 models he has shared. I would also like to thank
Z.~Bajnok,
E.V.~Frenkel,
M.~Gaberdiel
and A.~Kent
for helpful conversations, comments and criticism at various stages.
This work was supported by a fellowship from St.\ John's College,
Cambridge.
Some of the calculations were performed using the algebraic
manipulation programme REDUCE, and some were performed
on computers supplied under SERC `Computational Science
Initiative' Grant GR/H57585.



\end{document}